\documentclass[aps,prb,showpacs,floatfix]{revtex4}
\usepackage{graphicx}

\begin{document}
\title{Dark and Bright Excitonic States in Nitride Quantum Dots}

\author
{Anjana Bagga$^{\star}$, P. K. Chattopadhyay$^{\dag}$ and Subhasis Ghosh$^{\star} $}

\affiliation{$ ^{\star}$School of Physical Sciences, Jawaharlal Nehru University, New Delhi 110067\\
             $ ^{\dag}$Department of Physics, Maharshi Dayanand University, Rohtak}

\begin{abstract}
Formation of excitonic states in quantum dots, of nitride based III-V semiconductors GaN and AlN, including 
coulombic interaction, exchange interaction and dielectric effects are investigated. Dark exciton formation is found to occur for both GaN quantum dots(QDs) with wurtzite structure having positive crystal field splitting and GaN and AlN QDs with zinc-blende structure having zero crystal field splitting. The transition from dark to bright exciton occurs between radii range 20\mbox{\AA}$\sim$40\mbox{\AA} depending on the amount of dielectric mismatch between the dot and the surroundings. In wurtzite AlN QDs with negative crystal field splitting, the splitting between the dark and bright excitonic states is very small and vanishes at about $15$\mbox{\AA}.
\end{abstract}

\maketitle

\section{ Introduction}
\label{intro}

        A very interesting feature of semiconductor quantum dots(QDs) is the red shift of emission peaks with respect to absorption spectra and its size dependence\cite{PDJC93}$^-$\cite{UB97}. As the radius increases the red shift decreases and disappears beyond a certain radius. Such red shifts in QDs have been measured in Si\cite{PDJC93}, CdSe\cite{MN95}$^,$\cite{ALE96}, InAs\cite{UB97} and become zero at about $50$\mbox{\AA} for CdSe and at about $40$\mbox{\AA} for InAs. The theoretical explaination of the phenomenon is based on the formation of dark exciton states where the top of the valence band is either an optically passive P state\cite{GBG90}$^,$\cite{JL00} or the electron and the hole are in a triplet state\cite{ALE96}. Absorption of a photon and subsequent formation of an exciton can then take place only from an optically active state lying deeper in valence band. Such an exciton is termed dark as it cannot decay by a direct dipole transition to the top of the valence band. The decay eventually takes place with the help of phonons yielding red shifted photons.

The electronic energy levels of GaN, InN and AlN QDs without the electron hole(e-h) interaction have recently been investigated in detail by us\cite{AB03}. In this paper we study the excitonic states in nitrides as a function of the sign of the crystal field splitting $\Delta_{cr}$. The excitonic energy levels are found to change drastically as one goes from wurtzite GaN(+ve $\Delta_{cr}$)to zinc-blende GaN(zero $\Delta_{cr}$ ) and further to wurtzite AlN(-ve $\Delta_{cr}$). We examine the formation of dark and bright excitons in GaN and AlN QDs by considering the e-h coulomb interaction, e-h exchange interaction and the dielectric mismatch effects using the multiband $\bf{k.p}$ method. The suitability of the method for this type of problems is discussed in Ref.3 and 6. The nitride based QDs are interesting because (i) unlike GaAs and most other 
III-V semiconductors which have zinc-blende structure, III-V nitrides have both zinc-blende and wurtzite structure; (ii) the crystal field splitting is negative for wurtzite AlN(w-AlN) but is positive for wurtzite GaN(w-GaN); (iii) the  possibility of using the inbuilt electric field due to giant piezoelectric effect to generate entangled exciton-exciton states required for   quantum computers\cite{WDO00}. Since the crystal field effects are absent in the zinc-blende structures, one is able to study and compare spectra of systems with positive, zero and negative crystal fields. In Sec.II the selection rules for the optical transitions in QDs are given and the concept of dark exciton is introduced. Sec.III gives the theoretical framework used in the calculation. The results obtained and their discussion will be found in Sec.IV. This is followed by conclusions in Sec.V. 

\section{Selection Rules for Optical Transitions in QDs}
\label{Select}
The probabilities of the optical transition at band edges in QDs are governed by the matrix element of the operator $e\widehat{\bf{p}}$ between  the states at the valence band top and  the bottom of the conduction band, where $e\widehat{\bf{p}}$ is a momentum operator. These states incorporate both the effects of crystal structure and confinement. The electron wavefunction, $1S_{e}$, at the bottom of the conduction band can be written as \cite{ALE92}  
\begin{equation}
\label{h1}
\psi_{e}\left(\bf{r}\right)
 =j_{0}(k_{1}^{0}r)Y_{00}\left(\theta,\phi\right)\left|S\alpha\right\rangle,
\end{equation}
\noindent  where $ \left|S\alpha\right\rangle$ are the Bloch functions of the conduction band,  $\alpha$ is the projection of the electron spin, $Y_{00}$ is the spherical harmonics
and $j_{0}(k_{1}^{0}r)$ is the spherical bessel function.
   Optical transitions from this state are possible only to hole states in the valence band which have the S-state as a component because 
{\footnotesize $ \int Y_{Lm}Y_{00} d\Omega=\delta_{L,o}\delta_{m,o}$.} The optical transition
  probability from the bottom of the conduction band to a hole state in the valence band is given by \cite{ALE92}
$P_{\alpha\beta}=\left|\int dr  r^2 j_{0}(k_{1}^{0}r)
\left(cR_o(r)\right)\right|^{2}
\left|\left\langle u_{\mu}\beta\left|e\widehat{\bf{p}}\right|S\alpha\right\rangle\right|^{2}$,
where c is the component of the S-state $R_{o}(r)$ in the hole wavefunction,
 $\beta$ is the spin projection of the removed electron in the hole state and $u_{\mu}$ are the Bloch wavefunctions of the hole states:
{\footnotesize $u_{\pm}=\left|1,\pm1\right\rangle, u_{o}=\left|1,0\right\rangle$.}
$P_{\alpha\beta}$ is zero if $\alpha\neq\beta$. Optical transitions from the bottom of the conduction band to the top of the valence band is not possible if either    
    (1) The hole state does not have a component of the S-state (i.e if c=0), or
    (2) The spin projections of the removed electron in the hole state and the electron in the conduction band are  not equal, $\alpha\neq\beta$. Since the spin projection of the hole state is obtained by flipping the spin projection of the electron removed, it follows that optical transitions are not possible if the spins of the electron and the hole are parallel i.e if they are in a triplet state. Thus, if  the top of the valence band is a P-state, or the electron at  bottom of the conduction band and the hole at the top of the valence band are in a triplet state, a {\sl dark exciton} is formed. 
    
\section{Excitonic states in spherical quantum dots}
\label{state}
The exciton states are obtained from the Hamiltonian
\begin{equation}
\label{h3}	
H=H_{e}+H_{h}+H_{so}+V_{e-h}+V_{Pol-s}+V_{Pol-eh}	
\end{equation}
where $H_{e}, H_{h}, H_{so}, V_{e-h}$ are the electronic Hamiltonian, the hole Hamiltonian, the spin-orbit interaction and the e-h coulomb interaction respectively. The last two terms represent the surface polarization energies arising due to the difference in the dielectric constants between the semiconductor quantum dot and the surrounding medium\cite{TT93}. $V_{Pol-s}$ is the self energy of the electron and hole due to their image charges and $V_{Pol-eh}$ is the mutual interaction energy between the electron and hole via image charges. The excitonic states obtained from Eq.(\ref{h3}) will be split further due to the exchange interaction\cite{ALE96} $H_{ex}$ between the electron and hole.  

      $H_{e}$ is effectively the Hamiltonian of a "free" electron with effective mass $m_{e}^\ast$ confined to a spherical dot of radius R. The lowest eigenfunction of $H_{e}$
is given by Eq. (\ref{h1}). Since an electron created in a higher state of the conduction band will quickly cascade down to the lowest state, this state is the state of interest in exciton calculation. The hole Hamiltonian $H_{h}$ and the spin-orbit interaction $H_{so}$, are given in Refs. 6, 7 for wurtzite structures in the effective mass theory. The basis used is the direct product of Bloch wavefunctions at the valence band top with angular momentum $I=1$ and the spin eigenstates ($\left|I,I_{z}\right\rangle \left|S,S_{z}\right\rangle$):
 
{\footnotesize
\begin{equation}
\label{h4}
\begin{array}{ccc}
 \left|u_{1}\right\rangle=	\left|1,1\right\rangle \left|\frac{1}{2},      \frac{1}{2}\right\rangle, & \left|u_{2}\right\rangle=\left|1,0\right\rangle \left|\frac{1}{2},\frac{1}{2}\right\rangle, & \left|u_{3}\right\rangle=	\left|1,-1\right\rangle \left|\frac{1}{2},\frac{1}{2}\right\rangle \\
\left|u_{4}\right\rangle=	\left|1,1\right\rangle \left|\frac{1}{2},-\frac{1}{2}\right\rangle, & \left|u_{5}\right\rangle=	\left|1,0\right\rangle \left|\frac{1}{2},-\frac{1}{2}\right\rangle, &    \left|u_{6}\right\rangle=	\left|1,-1\right\rangle \left|\frac{1}{2},-\frac{1}{2}\right\rangle
\end{array}
\end{equation} }
   where $\left|1,1\right\rangle  =-\frac{1}{\sqrt{2}}\left|X+\iota Y\right\rangle,
\left|1,0\right\rangle  = \left|Z\right\rangle,
\left|1,-1\right\rangle = \frac{1}{\sqrt{2}}\left|X-\iota Y\right\rangle$. The hole Hamiltonian in this basis with the energy reference as the top of the valence band is given by  
{\footnotesize
\begin{equation}
\label{h5}	
H_{h}+H_{so}=\frac{1}{2m_{o}}
\left(\begin{array}{cccccc}
 P_{1}                  & S                      & -T                    &
  0                     & 0                      & 0                    \\
  S^{*}                 & P_{3}+2m_{o}\lambda    & -S                    & 
 -2\sqrt{2}m_{o}\lambda & 0                      & 0                    \\                   
 -T^{*}                 &-S^{*}                  & P_{1}+4m_{o}\lambda   &
  0                     & -2\sqrt{2}m_{o}\lambda & 0                    \\
  0                     & -2\sqrt{2}m_{o}\lambda & 0                     &
 P_{1}+4m_{o}\lambda    & S                      & -T                    \\                    0                      & 0                      &-2\sqrt{2}m_{o}\lambda & 
 S^{*}                  & P_{3}+2m_{o}\lambda    & -S                    \\
 0                      & 0                      & 0                     &
 -T^{*}                 & -S^{*}                 & P_{1}                                   
\end{array} \right)
\end{equation}   
}
where $\lambda=\frac{\Delta_{so}}{3}$, $\Delta_{so}$ being the splitting at the top of the  valence band $\Gamma$ due to the spin orbit interaction. 
$P_{1}=\gamma_{1}p^{2}-\sqrt{\frac{2}{3}}\gamma_{2}P_{o}^{\left(2\right)}$,   
$P_{3}=\gamma_{1}^{\prime}p^{2}+\sqrt{\frac{2}{3}}\gamma_{2}^{\prime}
        P_{o}^{\left(2\right)}+2m_{o}\Delta_{cr}$,   
$T    =\eta P_{-2}^{\left(2\right)}+\delta P_{2}^{\left(2\right)}$, 
$T^{*}=\eta P_{2}^{\left(2\right)}+\delta P_{-2}^{\left(2\right)}$,    
$S    =\Lambda p_{o}P_{-1}^{\left(1\right)}+
              \sqrt{2}\gamma_{3}^{\prime}P_{-1}^{\left(2\right)}$,   
$S^{*}= -\Lambda p_{o}P_{1}^{\left(2\right)}
        -\sqrt{2}\gamma_{3}^{\prime}P_{1}^{\left(2\right)}$
where $P^{\left(2\right)}$ and $P ^{\left(1\right)}$ are second and first rank spherical tensors formed out of the components $p_{x}, p_{y}, p_{z}$ and $\Delta_{cr}$ is the crystal field splitting energy. The parameters $\gamma_{1},
\gamma_{2},\gamma_{1}^{\prime},\gamma_{2}^{\prime} $ etc. are related to effective mass parameters $L, M, N \cdots$
by the relations\cite{JL00} 
$\gamma_{1}=\frac{1}{3} \left( L+M+N\right)$, $ \gamma_{2}=\frac{1}{3}\left(L+M+N \right)$
, $\gamma_{3}=\frac{1}{6}R $, 
$\gamma_{1}^{\prime}=\frac{1}{3}\left(T+2S\right)$,  $\gamma_{2}^{\prime}=\frac{1}{6}\left(T-S\right)$, $\gamma_{3}^{\prime}=\frac{1}{6}Q$,  
$\eta =\frac{1}{6}\left(L-M+R\right)$, $\delta=\frac{1}{6}\left(L-M-R\right)$.    
 The effective mass parameters $L,M,N...S,T$ have been taken from Ref.7.
While the basic structure in wurtzite is hexagonal close packed(hcp), the basic
structure in zinc-blende is face centered cubic(fcc). In zinc-blende structure crystal field
effects are absent and the transition from the Hamiltonian $H_{o}+H_{so}$ in
the wurtzite case to the corresponding one in the zinc-blende is easily
obtained by putting $\Delta_{cr}=0,\Lambda=0, R=Q=N, S=M, T=L$. Baldereschi and Lipari \cite{JBX96}$^,$\cite{AB73} have shown that for the zinc-blende structures the spherical symmetry approximation for the Hamiltonian is a good approximation.  

 The eigenfunction of the Hamiltonian given in Eq. {\ref{h5}} can be written as
\begin{equation}
\label{h6}
           \Psi_{m+ \frac{1}{2}}=\sum_{\ell,n}C_{n,\ell}j_{\ell}\left(k^\ell_n r\right)
\left(\begin{array}{c}
       a_{n,\ell}Y^{m-1}_\ell\left(\theta,\phi\right)\\
       b_{n,\ell}Y^m_\ell\left(\theta,\phi\right)\\
       d_{n,\ell}Y^{m+1}_\ell\left(\theta,\phi\right)\\
       a^\prime_{n,\ell}Y^m_\ell\left(\theta,\phi\right)\\
       b^\prime_{n,\ell}Y^{m+1}_\ell\left(\theta,\phi\right)\\
       d^\prime_{n,\ell}Y^{m+2}_\ell\left(\theta,\phi\right)
\end{array} \right) 
\end{equation}
In Eq. (\ref{h6}), $j_{\ell}\left(x\right)$ is the spherical Bessel function,                $k^\ell_n=\frac{\alpha^\ell_n}{R}$, where $\alpha^\ell_n$  is the $n^{th}$ zero of $j_{\ell}\left(x\right)$, R is the dot radius and $C_{n,\ell}$ is an overall normalization constant given by 
$C_{n,\ell}=\frac{\sqrt{2}}{R^{\frac{3}{2}}}\frac{1}{j_{\ell+1}\left(\alpha^\ell_n\right)}$ . It should be noted that the $\ell_{z}$-values in the last three terms of the column matrix, given
in Eq. (\ref{h6}), are one more than the corresponding values in the first three, because the former are associated with spin states with $ S_{z}=\frac{1}{2}$  while the later are associated with $ S_{z}=-\frac{1}{2}$. Each eigenfunction is characterized by a definite value of $M=m+ \frac{1}{2}=\ell_{z}+I_{z}+S_{z}$ where $\ell_{z}, I_{z}, S_{z}$ represent the z-components of the spherical harmonics, Bloch wavefunctions at the valence band top, and the spin part respectively. The energy eigenvalues and wavefunctions for the hole states are obtained from the solution of the Schrodinger equation
\begin{equation}
\label{h7}	
 H \Psi_{m+\frac{1}{2}}=E \Psi_{m+\frac{1}{2}}
\end{equation}
with $H$ given by Eq. (\ref{h5}) and $\Psi_{m+\frac{1}{2}}$ given by Eq. (\ref{h6}).

~~~~~~The coulomb interaction between the electron and the hole $V_{eh}$ is given by
$$V_{eh}=-\frac{e^2}{r_{eh}}$$
where $e(h)$ refers to the electron (hole). If we take the excitonic Bohr radius in the medium $a_{ex}=\hbar^2 \epsilon_d/m^*_ee^2$ and Rydberg $R_{ex}=m^*_ee^4/2\hbar^2 \epsilon^2_r$ ($m^*_e$ is the effective mass of the electron in units of the free electron mass $m_0$, and $\epsilon_d$ is the dielectric constant of the dot material) as the units of length and energy respectively, $V_{eh}$ becomes 
\begin{equation}
V_{eh} = - {2 \over r_{eh}},
\end{equation}
The exciton wavefunction can be expanded in terms of the electron and hole wavefunctions as
\begin{equation}
\label{h8}
\psi_{ex} = \sum\limits_{i,j} C_{ij} \psi_{ei} (\vec{r}_e) \psi_{hj} (\vec{r}_h),
\end{equation}
where $\psi_{ei} (\vec{r}_{e})$ is the eigenstate of the electron of the conduction band and $\psi_{hj} (\vec{r}_{h})$ is the hole eigenstate obtained from Eq. {\ref{h7}}. The exciton energy can be obtained from the secular equation
\begin{equation}
|(E_{n_e,l_e} + E_{m_h,l_h} - E) \delta_{ik} \delta_{jl}+ V_{ij,kl}| = 0
\end{equation}
where\cite{JL00}
$$ V_{ij,kl}=\left\langle \psi_{ei} (\vec{r}_e) \psi_{hj} (\vec{r}_h)\left|\frac{-2}{r_{eh}}\right|\psi_{ek} (\vec{r}_e) \psi_{hl} (\vec{r}_h)\right\rangle $$

The effects due to the difference in the dielectric constant between the semiconductor quantum dot and the surrounding medium in the excitonic Hamiltonian are given by $V_{Pol-s}$ and $V_{Pol-eh}$. These two terms represent the surface polarization energies and are given by \cite{TT93}
\begin{equation}
\label{h9}
\begin{array}{cccc}
V_{Pol-s}&=&& \frac{e^{2}}{2R}\sum\limits^{\infty}_{n=0}\alpha_{n}\left[\left[\frac{r_{e}}{R}\right]^{2n}+\left[\frac{r_{h}}{R}\right]^{2n}\right]\\
V_{Pol-eh}&=&-& \frac{e^{2}}{2R}\sum\limits^{\infty}_{n=0}\alpha_{n}\left[\frac{r_{e}r_{h}}{R^{2}}\right]^{n}P_{n}\left(cos\theta_{eh}\right)
\end{array}
\end{equation}
where $P_{n}$ is the Legendre polynomial of nth order and $\theta_{eh}$ is the angle between
$r_{e}$ and $r_{h}$. The dielectric constants of the semiconducting material of the dot and the surrounding medium are denoted by $\epsilon_{d}$ and $\epsilon_{s}$ respectively and $\alpha_{n}$ is defined by $\alpha_{n}=\frac{(n+1)(\epsilon-1)}{\epsilon_{s}(n\epsilon+n+1)}$
with $\epsilon=\frac{\epsilon_{d}}{\epsilon_{s}}$. Both $V_{Pol-s}$ and $V_{Pol-eh}$ become zero when either the radius of the dot $R\rightarrow\infty$ or $\epsilon=\frac{\epsilon_{d}}{\epsilon_{s}}=1 (\alpha_{n}=0)$. Both the cases correspond to no dielectric mismatch between the system and the surroundings.

In addition to the coulomb interaction there is the exchange interaction    \cite{ALE96}$^,$\cite{TT93} between the electron and the hole which splits the active exciton states described above and is given by 
\begin{equation}
\label{h10}	
H_{ex}=\left(-\frac{2}{3}\right)\epsilon_{ex}(a_{o})^{3}\delta(\vec{r_{e}}-\vec{r_{h}})
\bf{\sigma} \bf{J}
\end{equation}
where $\epsilon_{ex}$ is the exchange strength constant, $a_{o}$ the lattice constant,
$\bf{\sigma}$ the Pauli spin matrices representing the electron spin and $\bf{J}$
is the hole spin matrix. The exchange interaction was diagonalized in the eight dimensional basis\cite{ALE96} obtained by taking the direct product of the hole S-states with $M=\pm\frac{3}{2}$ and
$M=\pm\frac{1}{2}$ and the electron spin states $\pm\frac{1}{2}$.

\section{Results and Discussion}
\label{model}
    The transition probability between the hole state at the top of the valence band and the electron S-state at the bottom of the conduction band is zero if either (1) the hole state is a P state and not a S-state or (2) if the hole at the top of the valence band and the electron at the bottom of the conduction band are in a triplet state. Hence in the above two cases a dark exciton is formed between the electron and the hole. In both the cases the excitation takes place from a hole state, having an S-state component, lying deeper in the valence band, to the electron S-state at the bottom of the conduction  band. This is a transition to a singlet state between the hole and the electron which has higher energy than the triplet state. De-excitation takes place eventually, with the help of phonons, to the P state at the top of the valence band in the first case and to the triplet state and finally to the S-state of the valence band in the second case. This gives rise to the red shift of the emission spectra relative to the absorption spectra in both the cases. In view of the different mechanisms involved, the discussion is divided into two subsections. In the first subsection we investigate the nature of the state at the top of the valence band i.e whether the state is an optically passive P-state or an optically active S-state. This is done first without including the electron hole coulomb interaction and then with the inclusion of excitonic effects. The effects of dielectric mismatch between the QD and the surrounding medium have also been considered. In the next subsection we study the effect of the exchange interaction between the electron at the bottom of the conduction band and the hole at the top of the valence band on the excitonic states. The exchange interaction gives rise to a splitting between the triplet and the singlet state with the triplet state lying lower in energy. The red shift, which is the singlet-triplet energy difference, is a function of the radius of the QD. It decreases with the increase of radius and vanishes beyond a certain radius. As observed in our previous work(Ref.7), the crystal field plays a crucial role in determining the nature of the state at the top of the valence band and hence the red shift. The effect of the sign of the crystal field splitting $\Delta_{cr}$ on the red shift has been investigated by comparing the results for wurtzite GaN, zinc-blende GaN and wurtzite AlN QD's having positive, zero and negative crystal field splitting $\Delta_{cr}$ respectively.

\subsection{Coulombic interaction and the dielectric mismatch effects}
\label{coulomb}
The calculated hole spectra for GaN and AlN including excitonic effects  as a function of the dot radius R are shown in Figs.1-3. The eigenvalues and the eigenfunctions of $H_{h}+H_{so}$ have been investigated in detail in our previous work (Ref.7) but here we focus on the trends of the first optically passive P state and the first optically active S state as a
function of the dot radius R. For comparison Figs.1(a), 2(a) and 3(a) give the energies of the hole states without the e-h coulomb interaction for w-GaN with +ve $\Delta_{cr}$,
 zinc-blende GaN(z-GaN) with zero $\Delta_{cr}$ and w-AlN with -ve $\Delta_{cr}$ respectively. In Figs.1(b), 2(b) and 3(b) are given the first optically passive and the first optically active states of the same materials with the e-h coulomb interaction included. Both the cases, when the dielectric effects are present and absent have been considered. The energies are in the units of   
$ \epsilon_{o}=\frac{\gamma_{1}}{2m_{o}}\left(\frac{\hbar}{R}\right)^{2}$.  The following observations are made in three different cases:

\noindent {\bf 1. Wurtzite GaN ($\Delta_{cr}$ is positive):} Fig.1(a) gives the hole eigenvalues in the absence of e-h interaction. The states are labelled as $\left|P_{x}\uparrow\right\rangle$, $\left|S_{z}\downarrow\right\rangle$ etc. Since the wurtzite Hamiltonian is axially symmetric only the z-component of the total angular momentum $\bf{F^{\prime}}$=$\bf{\ell}+\bf{I}+\bf{S}$
is conserved where $\bf{\ell}$ is the orbital angular momentum, $\bf{I}$(=1) is angular momentum of the Bloch wavefunction at the valence band top and $\bf{S}$ is the spin of the hole. In  labelling of the states the capital letters (S, P etc.) represent the dominant $\bf{\ell}$ present and the subscripts indicate whether the states are X-, Y-like or Z-like, the arrow indicating the spin state. We observe the following trends for the lowest optically passive P state and the first optically active S state :(1) The ground state is $\left|P_{x}\uparrow\right\rangle$ rather than an S-state, with the z-component of total angular momentum $M=\frac{1}{2}$.
(2) As seen from Table.I (which gives the probabilities of the different components of angular momentum in the  ground and first excited states), the ground state $\left|P_{x}\uparrow\right\rangle$ has about $50\%$ probability in the state with $\ell=1, I_{z}=+1, S_{z}=+\frac{1}{2}$ and $49\%$ probability in the state with $\ell=1, I_{z}=-1, S_{z}=+\frac{1}{2}$ for small R. As R increases the energy
of the state increases, the probability in the state with $I_{z}=+1$ increases and the probability in the state with $I_{z}=-1$ decreases. (3) The first optically active state for the system is $\left|S_{x}\uparrow\right\rangle$ with $M=\frac{3}{2}$. This state is relatively flat as a function of R and has a large probability only in the state with $\ell=0, I_{z}=+1, S_{z}=+\frac{1}{2}$.\\
  Thus when the e-h coulomb interaction is not taken into account, the system is predicted to have dark excitons for even very large values of R.
  
{\it Inclusion of the Coulomb interaction between the electron and the hole:}  The situation changes significantly when the e-h coulomb interaction is taken into account. Fig.1(b) with $\epsilon=\frac{\epsilon_{d}}{\epsilon_{s}}$ ($\epsilon_{d}$= dielectric constant inside the dot, $\epsilon_{s}$= dielectric constant of the surrounding medium) shows that for $\epsilon=1$(no dielectric mismatch), with the inclusion of the e-h coulomb interaction, the energies of both the passive  $\left|P_{x}\uparrow\right\rangle$ state and the first active  excited state $\left|S_{x}\uparrow\right\rangle$ have a downward slope as a function of R. With the energy of $\left|S_{x}\uparrow\right\rangle$ decreasing faster, at about $R \sim 50$\mbox{\AA} the levels cross and the active $\left|S_{x}\uparrow\right\rangle$
state becomes the ground state. Thus there will be a transition from dark excitonic state to bright excitonic state for w-GaN at about $R \sim 50$\mbox{\AA}.   

 At very low radii, the kinetic energy due to confinement is very high. As the radius increases, the kinetic energy due to confinement decreases and the relative contribution of the coulomb energy to the total energy increases. Since the coulomb interaction is stronger in the S-state than in  the P state, the decrease in energy of the S-state is more than that for the P-state, the gap between $\left|P_{x}\uparrow\right\rangle$ and
 $\left|S_{x}\uparrow\right\rangle$ decreases and the levels eventually cross. Figs.5(a) and 6(a) give the energy difference between the $\left|S_{x}\uparrow\right\rangle$ and $\left|P_{x}\uparrow\right\rangle$ states as a function of R for w-GaN and z-GaN respectively.    

{\it Effect of dielectric mismatch between the QD and the surroundings : } The
effects of dielectric mismatch between the dot and the surrounding medium can be seen from 
Fig.1(b) by comparing the corresponding curves for $\epsilon=5$(dielectric mismatch present)
and $\epsilon=1$(no dielectric mismatch). The dielectric mismatch effects are greatly reduced because of large cancellations of contributions from the self energies of the electrons and holes due to their image charges (represented by $V_{Pol-s}$ in Eq. (\ref{h9})) and the mutual interaction between the electron and the hole via image charges $(V_{Pol-eh}$ in Eq. (\ref{h9})). The contributions are of opposite sign but comparable in magnitude\cite{VAF02}, the net effect being an increase in energy. The inclusion of surface polarization effects causes the cross-over between $\left|S_{x}\uparrow\right\rangle$ and $\left|P_{x}\uparrow\right\rangle$ states to occur at lower radii, the amount of shift depending on the amount of dielectric mismatch $\epsilon$. For $\epsilon=5$, the cross-over occurs at about $R \sim 30$\mbox{\AA}. This happens because the energy of the P state is raised more due to surface polarization effects than that of the S-state.
The center of the dot r=0 is the position of greatest dielectric stabilization\cite{LEB83}. Since $\left\langle r^{2}\right\rangle$ of the charge distribution is higher for the P-state than for the S-state, the increase in energy due to surface polarization effect term $V_{Pol-s}$ is higher for the P-state than for the S-state. $V_{Pol-eh}$ representing the coulombic interaction is stronger in the S state than in the P state. Since this term is negative, it also makes the expectation value of $V_{Pol-s} + V_{Pol-eh}$ higher for the P-state than for the S-state. Figs.5(a) and 6(a) for w-GaN and z-GaN respectively give the energy difference in meV as a function of R with the dielectric mismatch effects included.

\noindent {\bf 2. Zinc-Blende GaN ($\Delta_{cr}$ is zero):} Fig.2(a) gives the hole eigenvalues in the absence of e-h interaction. The states are labelled as $S_{\frac{3}{2}}$, $P_{\frac{1}{2}}$ etc where capital letters correspond to the lowest $\vec{\ell}$ present and the subscripts gives the total angular momentum $\bf{{F^{\prime}}}$. Unlike the wurtzite
case the total angular momentum $\vec{F^{\prime}}$ is conserved for the zinc-blende. We observe that the ground state is the optically passive $P_{\frac{3}{2}}$ rather than the optically active $S_{\frac{3}{2}}$ state. The ground state $P_{\frac{3}{2}}$ is degenerate with respect to its z-component $M=\pm\frac{3}{2}, \pm\frac{1}{2}$. It is also almost degenerate with the $P_{\frac{1}{2}}$ upto about $R \sim 10$\mbox{\AA} and slopes downwards as R increases. The first optically active state $S_{\frac{3}{2}}$ lies above $P_{\frac{3}{2}}$ and is almost degenerate with $S_{\frac{1}{2}}$ upto about $R \sim 10$\mbox{\AA} and slopes downwards as $R$ increases.

\noindent Thus as in the case of w-GaN, for z-GaN also the ground state is a dark excitonic state. When the coulomb interaction is included without the inclusion of dielectric mismatch effects (Fig.2(b) with $\epsilon=1$), the energies of both the states decrease faster with R with the $S_{\frac{3}{2}}$ state eventually crossing the $P_{\frac{3}{2}}$ and becoming the ground state at about $R \sim 35$\mbox{\AA}. As in the case of w-GaN, 
the inclusion of surface polarization effects shifts the transition from dark to bright excitonic state  to lower R (Fig.2(b) with $\epsilon=5$). For $\epsilon=5$, the crossover occurs at $R \sim 20$\mbox{\AA} as seen from Fig.2(b).

\noindent{\bf 3. Wurtzite AlN ($\Delta_{cr}$ is negative):} Fig.3(a) gives the hole eigenvalues in the absence of e-h interaction. The states are labelled in the same manner as for w-GaN. We observe that the active $\left|S_{z}\uparrow\right\rangle$ state with $M=\frac{1}{2}$ is the ground state beyond $R \sim 20$\mbox{\AA} in the absence of e-h interaction. This state is very flat throughout the radii range with high probability only for the state with ${\bf \ell}=0, I_{z}=0,$ and $ S_{z}=+\frac{1}{2}$ as seen from Table. I. After the inclusion of e-h coulomb interaction and the surface polarization effects, $\left|S_{z}\uparrow\right\rangle$ becomes the ground state for all radii as seen from Fig. 3(b). Therefore there will be no dark exciton in  wurtzite AlN QDs. 

    Thus it is found that
    (i) dark exciton is formed in GaN QDs with zinc-blende structures ($\Delta_{cr}$=0) and with wurtzite structures ($\Delta_{cr}$ positive) for radii smaller than $30 \sim 40$\mbox{\AA}. There is no dark exciton for AlN QDs with wurtzite structure ($\Delta_{cr}$ negative). The behaviour of zinc-blende AlN will be similar to zinc-blende GaN. 
    (ii) The first optically active S-state comes from $M=\pm\frac{3}{2}, \pm\frac{1}{2}$ for z-GaN QD, from $M=\pm\frac{3}{2}$ for w-GaN and from $M=\pm\frac{1}{2}$ for w-AlN.    

    The observed features of the first optically passive P-state and the first optically active S-state of the hole( eigenstates of $H_{h}+H_{so}$) can be explained by analyzing the Hamiltonian given in Eq. {\ref{h5}}. For wurtzite GaN QD structures $\Delta_{cr}$ is positive and the states  with $I_z=\pm1$ lie lower to the state with $I_z=0$. For $M=\ell_{z}+I_{z}+S_{z}=\frac{1}{2}$ with $I_{z}=\pm1$ and $S_{z}=\frac{1}{2}$, $\ell_{z}$ should be $\mp1$. The lowest $\vec{\ell}$ for which this can happen is $\vec{\ell}=1$. Hence for $M=\frac{1}{2}$, the lowest state is a P-state and not a S-state and is primarily made of $I_{z}=\pm1$. This also shows that the first optically active S-state comes from $M=\frac{3}{2}$ with the composition $\ell=0, \ell_z=0, I_z=+1, S_z=+\frac{1}{2}$. For QDs of w-AlN  with -ve $\Delta_{cr}$ the state with $I_z=0$ is lower than states with $I_z=\pm1$. For $M=\ell_{z}+I_{z}+S_{z}=\frac{1}{2}$ with $I_{z}=0$
and $S_{z}=\frac{1}{2}$, $\ell_{z}$ must be $0$. The lowest $\vec{\ell}$ for which this is possible is $\vec{\ell}=0$ and we have the S-state as the ground state. 

     For GaN QDs with the +ve $\Delta_{cr}$, in the absence of e-h coulomb interaction, the P-state with $M=\frac{1}{2}$ lies lower than the S-state with $M=\frac{3}{2}$ and slopes upwards. This can be explained by looking at the probabilities for various states  given in Table.1. It is found that for small R, the $\left|P_{x}\uparrow\right\rangle$ state has
$50\%$ probability in the state with
$I_{z}=+1, S_{z}=+\frac{1}{2}$ and $49\%$ probability in the state with 
$I_{z}=-1, S_{z}=+\frac{1}{2}$. As R increases the energy of the state increases, the probability in the state with $I_{z}=+1$ increases and that in the state with $I_{z}=-1$decreases. The observed trends follow from the submatrix $H^{\prime}$
formed with base states $\left|1,1\right\rangle \left|\frac{1}{2},      +\frac{1}{2}\right\rangle$ and $\left|1,-1\right\rangle \left|\frac{1}{2},      +\frac{1}{2}\right\rangle$ of the Hamiltonian given in Eq. (\ref{h5}) 
{\footnotesize
\begin{equation}
\label{hd6}	
H^{\prime}=\frac{\hbar^{2}}{2m_{o}R^{2}}
\left(\begin{array}{cccccc}
 P_{1}^{\prime}    & -T^{\prime}\\
 -T^{\prime*}      & P_{1}^{\prime}+\lambda^{\prime} R^{2}                          
\end{array} \right)
\end{equation} 
}
   where $\lambda^{\prime}=\frac{4m_{o}\lambda R^{2}}{\hbar^{2}}$. Since the graphs are in the units of $ \epsilon_{o}=\frac{\gamma_{1}}{2m_{o}}\left(\frac{\hbar}{R}\right)^{2}$
the ${(\frac{\hbar}{R})}^{2}$ dependence coming from $p^{2}$ term contained in $P_{1}$  
and T of Eq. (\ref{h5}) has been taken out. $P_{1}^{\prime} and -T^{\prime}$ are independent of R. At very small R, the  spin-orbit term involving $\lambda$ becomes negligible relative to momentum dependent terms $P_{1}$ due to confinement effects
and therefore $H^\prime_{11} \sim H^\prime_{22}$. Hence the state with $I_{z}=+1$ is equally probable as the state with $I_{z}=-1$. As the radius increases the confinement effect reduces and the spin orbit term $\lambda$ becomes more important. Hence 
the probability of the state with $I_{z}=+1$ in the ground state increases, that of the state with $I_{z}=-1$ decreases and the lowest eigenvalue slopes upwards. The 
$\left|S_{x}\uparrow\right\rangle$ state with $M=\frac{3}{2}$ has high probability only in the state with $I_{z}=+1, S_{z}=+\frac{1}{2}$ as seen from Table. 1 and is therefore primarily determined by $H^\prime_{11}$ which in the units of $ \epsilon_{o}$ is independent of R. This explains the relatively flat curve for the $\left|S_{x}\uparrow\right\rangle$
 state in Fig.2(a) and Ref.7. When the attractive coulomb attraction is added, the energy of the $\left|S_{x}\uparrow\right\rangle$ decreases faster than that of the $\left|P_{x}\uparrow\right\rangle$ since the coulomb interaction is stronger in the S state 
 than in P state. At about $R \sim 50$\mbox{\AA} the S state crosses the P state and becomes the ground state. Further inclusion of surface polarization effects shifts the above crossover to lower R as the increase in energy of the state due to surface polarization effects is more for the P state then for the S state.
 
  For QDs with zinc-blende structures, without the e-h interaction the $P_{\frac{3}{2}}$ state is degenerate with $P_{\frac{1}{2}}$ state upto $R \sim 10$\mbox{\AA} because at low R the contribution to energy of the spin-orbit interaction becomes negligible compared to the contributions from momentum dependent terms due to confinement. As the radius increases, the effect of spin orbit interaction becomes more important and the splitting between $P_{\frac{3}{2}}$ and $P_{\frac{1}{2}}$ increases. $P_{\frac{3}{2}}$ slopes downwards and $P_{\frac{1}{2}}$ slopes upwards. As mentioned above after the inclusion of couloumb interaction the  $S_{\frac{3}{2}}$
state crosses the $P_{\frac{3}{2}}$ state at $R \sim 30$\mbox{\AA}. This crossover shifts to lower R on further addition of dielectric mismatch effects, with the amount of shift depending on the amount of dielectric mismatch $\epsilon$, between the system and the surroundings. For $\epsilon=5$, the crossover occurs at $R \sim 20$\mbox{\AA} as seen from Fig.2(b).

\subsection{Inclusion of Exchange Interaction}
\label{exchange}
In addition to the coulomb interaction there is the exchange interaction    \cite{ALE96}$^,$\cite{TT93} between the electron and the hole which splits the active exciton states described above. We describe below the effect of exchange interaction in wurtzite GaN, zinc-blende GaN and wurtzite AlN QDs having positive, zero and negative crystal field splitting respectively.

\noindent {\bf 1. Zinc-Blende GaN ($\Delta_{cr}$ is zero):} A bright exciton is formed when an electron in an S-state in the valence band 
absorbs a photon and is lifted to the $1S_{e}$ state of the conduction band. For quantum dots with the zinc-blende structure the hole S-state has total angular momentum $\vec{F^{\prime}}=\frac{3}{2}$ and is four fold degenerate w.r.t its z-component $M=\pm\frac{3}{2}, \pm\frac{1}{2}$. The $1S_{e}$ state in the conduction band being doubly degenerate, in the absence of exchange interaction the exciton states in zinc-blende are eight fold degenerate.
 The exchange interaction splits this degenerate level into a lower(L) five fold degenerate 
 state with $\vec{F}=2 \left(\vec{F}=\vec{F^\prime}+\vec{S_{e}}\right)$ and $F_{z}=\pm2^L,\pm1^{L},0^{L}$ and an upper three fold degenerate level with 
$\vec{F}=1$ and $F_{z}=\pm1^{U}, 0^{U}$. 
 The exchange interaction Eq. (\ref{h10}) was diagonalized in the eight dimensional basis obtained by taking the direct product of the hole S-states with $M=\pm\frac{3}{2}$ and
 $M=\pm\frac{1}{2}$ and the electron spin states $\uparrow$ and $\downarrow$. In Fig.2(c)
 the energies of the two states with $\vec{F}=2$ and $\vec{F}=1$ in meV for z-GaN are plotted as a function of R. The energy without the exchange interaction is taken as zero. The $\vec{F}=2$ state is a triplet state. (The only way a state with total angular momentum 2 can be obtained from the electron and hole in the S state is by coupling the spins of the hole and electron to 1 and then coupling  it with I=1). Hence it is optically passive and forms a dark exciton. The upper state with F=1 is a singlet state. It is optically active and forms a bright exciton. The amount of red shift which is the splitting between the singlet and triplet states at any given value of R, decreases as R increases and becomes almost zero at $R \sim 50$\mbox{\AA} as seen from Fig.6(b).
 
\noindent {\bf 2. Wurtzite GaN ($\Delta_{cr}$ is positive):}  For these structures, the S state with $M$=$\pm \frac{3}{2}$ has a lower energy than the S state with $M$=$\pm \frac{1}{2}$. The energies of the exciton states in meV after including the exchange interaction for w-GaN as a function of R are shown in Fig.1(c). In this case the midpoint of the exciton energies with $M$=$\pm \frac{3}{2}$ and $M$=$\pm \frac{1}{2}$ is taken as the reference energy (zero of energy). The ground state with $F_{z}$=$\pm 2$ can be obtained 
 by coupling a hole state with $M$=$\pm \frac{3}{2}$ with the electron state with $(S_{e})_{z}$=$\pm \frac{1}{2}$. Since these combinations form triplet states, the ground state with $F_{z}$=$\pm 2$ is optically passive and forms a dark exciton. The first excited state with $F_{z}$=$\pm 1^{L}$ can be obtained either by coupling $M$=$\pm\frac{3}{2}$, $(S_{e})_{z}$=$\mp\frac{1}{2}$  or alternatively with $M$=$\pm\frac{1}{2}$, $(S_{e})_{z}$=$\pm\frac{1}{2}$. The state $\pm1^{L}$ will be a mixture of singlet and triplet states and will be optically active. The singlet triplet splitting as a function of R, in the presence and absence of dielectric mismatch effects, is given in Fig.5(b). It is observed that this splitting almost vanishes beyond $R \sim 40$\mbox{\AA} and is hardly affected by the dielectric mismatch effects.

\noindent {\bf 3. Wurtzite AlN ($\Delta_{cr}$ is negative):} For these structures, the S-state with $M$=$\pm\frac{1}{2}$ lies lower than the S-state with $M$=$\pm\frac{3}{2}$ because $\Delta_{cr}$ is negative. This causes a complete reordering of the excitonic levels in w-AlN as compared to w-GaN (see Fig.1(c) and Fig.4). The optically passive $F_{z}=0^{L}$ state is the ground state in w-AlN which is almost degenerate with the optically active
$\pm1^{U}$ states. This can be understood as follows:  The state $M$=$\pm\frac{1}{2}$ (with
 $\ell_{z}=0, I_{z}=0$ and  $S_{z}=\pm \frac{1}{2}$) can couple with the $(Se)_z=\pm\frac{1}{2}$ to give total z-component $F_{z}$=$0,0,\pm1$. These four states are 
 degenerate in the absence of exchange interaction. The exchange interaction couples the two states with $F_{z}=0(M=\pm\frac{1}{2},(Se)_{z}=\mp\frac{1}{2})$ and the degeneracy between the two is removed. But the splitting between the two is small compared to $\Delta_{cr}$.
The $F_{z}=\pm1$ state can be obtained either with $M=\pm\frac{1}{2},(Se)_{z}=\pm\frac{1}{2}$
or with $M=\pm\frac{3}{2},(Se)_{z}=\mp\frac{1}{2}$. In the absence of exchange interaction the excitonic state with $M=\pm\frac{3}{2},(Se)_{z}=\mp\frac{1}{2}$ lies above the excitonic state with $M=\pm\frac{1}{2},(Se)_{z}=\pm\frac{1}{2}$. The splitting between these states is large due to the presence of the large crystal field in w-AlN. The exchange interaction which couples these two states shifts their energies only slightly because the coupling between these states is weak compared to $\Delta_{cr}$. The lower $\pm1$ state is predominantly made of the state with $M=\pm\frac{1}{2},(Se)_{z}=\pm\frac{1}{2}$. Hence the states $0^{L},0^{U},\pm1^{U}$ remain close to one another even when the exchange interaction is switched on. The two $F_{z}=0$ states and the $\pm1$ states with $M=\pm\frac{1}{2}$ were degenerate in the absence of exchange interaction. Since both $0^{L} and \pm1^{U}$ decrease in energy when the exchange interaction is included, they come close to each other as seen from Fig.4. These states are also close together in w-GaN but in that case they lie above the $\pm2^{L}$ and $\pm1^{L}$ states because $\Delta_{cr}$ is positive(Fig.1(c)). The splitting between $0^{L}$ and $\pm1^{U}$ is further reduced in w-AlN compared to w-GaN because the enhancement factor $(\frac{a_{ex}}{R})^{3}$ for splitting in dots over the bulk value (see Refs. 14, 15), where $a_{ex}$ is the exciton bohr radius and R is the dot radius,
is smaller in w-AlN due to the smaller excitonic Bohr radius compared to w-GaN. The splitting between 
$0^{L}$ and $\pm1^{U}$ almost vanishes at $R \sim 15$\mbox{\AA}.

The red shift of the emission spectra with respect to the absorption spectra 
involving the excitation-deexcitation processes at the valence band top is shown schematically in Fig.7. Absorption takes place from the S-state which lies deeper than the P-state in the valence band to form an exciton in the singlet state. Deexcitation can take place from the exciton triplet state, reached by thermalization, either to the P state at the top of the valence band or to the original S-state lying deeper in the valence band with the help of phonons. In the first case, (1) of Fig.7, the stokes shift will be $\Delta E_{SP}+\Delta E_{ST}$ and in the second case, (2) of Fig.7, it will be $\Delta E_{ST}$ where $\Delta E_{SP}$ is the difference in energy between the exciton states formed with S and P hole states and $\Delta E_{ST}$ is the singlet triplet splitting. A third possibility, (3) of Fig.7, is direct deexcitation from the exciton singlet state to the P state at the top of the valence band, again with the help of phonons. In this case the Stokes shift will be $\Delta E_{SP}$. It can be seen from Figs.5 and 6 that
$\Delta E_{SP}>>\Delta E_{ST}$.

\section{Conclusions}
\label{conclude}

Dark and bright exciton formation have been studied in wurtzite GaN, zinc-blende GaN and wurtzite AlN QDs having positive, zero and negative crystal field splitting. It is found that\\ 
(1) For w-GaN(+ve $\Delta_{cr}$) QDs and z-GaN(zero $\Delta_{cr}$) QDs the optically passive P state forms the ground state of the hole. In the case of w-AlN(-ve $\Delta_{cr}$) QDs the optically active S-state is the ground state after 17\mbox{\AA}.\\
(2) For zinc-blende GaN QDs the first optically active S-state of the hole comes from $M=\pm\frac{3}{2}, \pm\frac{1}{2}$. For wurtzite GaN QDs with +ve $\Delta_{cr}$ the first optically active S-state corresponds to $M=\pm\frac{3}{2}$ and for $\Delta_{cr}$ negative the first optically active S state has $M=\pm\frac{1}{2}$. \\
(3) When the coulomb interaction between the electron and hole is added, the S state energy is lowered more than the P state energy. This results in the crossing between the active S-state and the passive P state and hence there is a transition from dark to bright excitonic state at about R $\sim$ 50\mbox{\AA} for w-GaN and at about R $\sim$ 35\mbox{\AA} for z-GaN QDs. For w-AlN QDs with negative crystal field splitting the active S state is the ground state throughout the radii range. \\
(4) Surface polarization effects due to the dielectric mismatch between the QD and the surroundings, shift the crossover to lower radii, the shift depending on the amount of dielectric mismatch. For $\epsilon=\frac{\epsilon_{d}}{\epsilon_{s}} = 5$ the transition from dark to bright excitonic state occurs at about R $\sim$ 30\mbox{\AA} for w-GaN and about R $\sim$ 20\mbox{\AA} for z-GaN QDs.\\
(5)The inclusion of the exchange interaction gives rise to splitting between the triplet state and the singlet state of the exciton with the triplet state having lower energy. In the case of z-GaN QDs the ground triplet state is characterized by total angular momentum $F=2$. For w-GaN only the z-component of the angular momentum M is a good quantum  number and $F_z=\pm2$ forms the ground state. The singlet triplet splitting decreases  with R and is almost zero at about $R \sim 50$\mbox{\AA}. For w-AlN with -ve $\Delta_{cr}$ there is a complete reordering of states relative to w-GaN case and the state with $F_z=0^L$ forms the ground state. The splitting between the passive $0^{L}$ and the active $\pm1^{U}$ state is very small and becomes almost zero at about $R \sim 15$\mbox{\AA}.

\newpage

\section*{Table Captions}
\noindent

\vspace*{.5cm}
\noindent
Table I :
The probabilities of different components for {$\bf M=\frac{1}{2},\frac{3}{2}$} in the low lying states, starting from the ground state, of wurtzite QDs
of AlN and GaN for radii $R=16$\mbox{\AA} and $R=76$\mbox{\AA}. For example in state 1 in GaN $(P,+1,\uparrow)=0.49$  indicates that the probability of the component with $\ell=1$
and $I=1$, $I_{z}=+1$ with spin up is 0.49. The table illustrates how the structure of the states changes with QD radius and material \cite{AB03}.  \\

\newpage

\section*{Figure Captions}
\noindent

\vspace*{.5cm}
\noindent
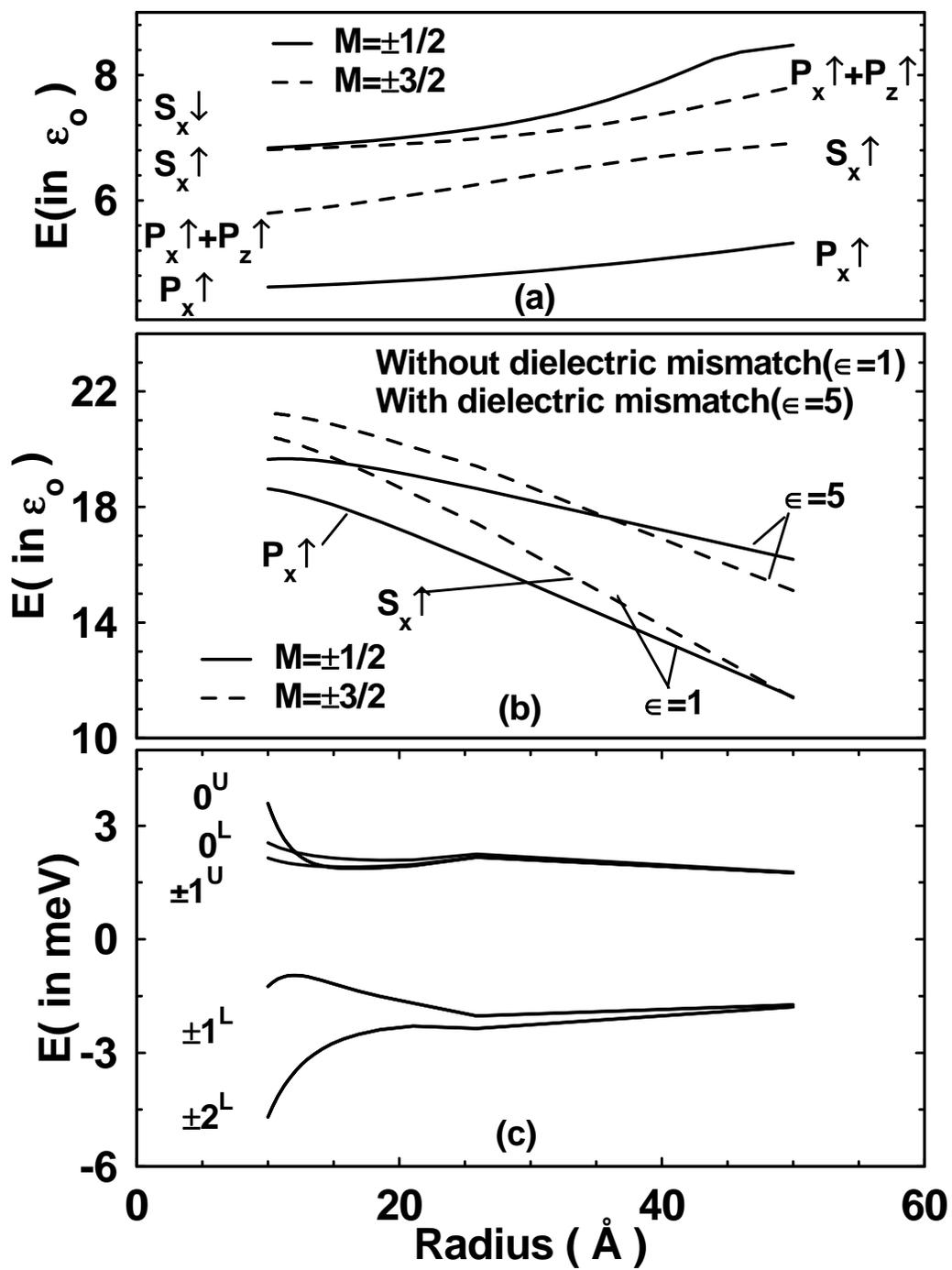
Fig.1:
The energy levels in units of $\epsilon_{0}$=$\frac{\gamma_{1}}{2m_{o}}\left(\frac{\hbar}{R}\right)^{2}$ for w-GaN QDs as a
function  R in (a) absence and (b) presence of e-h coulomb interaction(without exchange). Eigenvalues in the absence($\epsilon=1$) as well as presence($\epsilon=5$ of dielectric mismatch effects are also shown in (b).  The states are labelled as $S_{x}\downarrow, P_{x}\uparrow$ etc, where capital letters represent  the dominant $\bf{L}$ present and the subscripts for X-, Y-like or Z-like states, arrow indicating the spin state. (c) eigenvalues  including e-h coulomb and exchange interactions. The states are labelled with $F_{z}$ with superscripts standing for upper and lower state. The midpoint of the exciton energies with $M$=$\pm \frac{3}{2}$ and $M$=$\pm \frac{1}{2}$ is taken as the reference energy (zero of energy).
     
\vspace*{.5cm}
\noindent
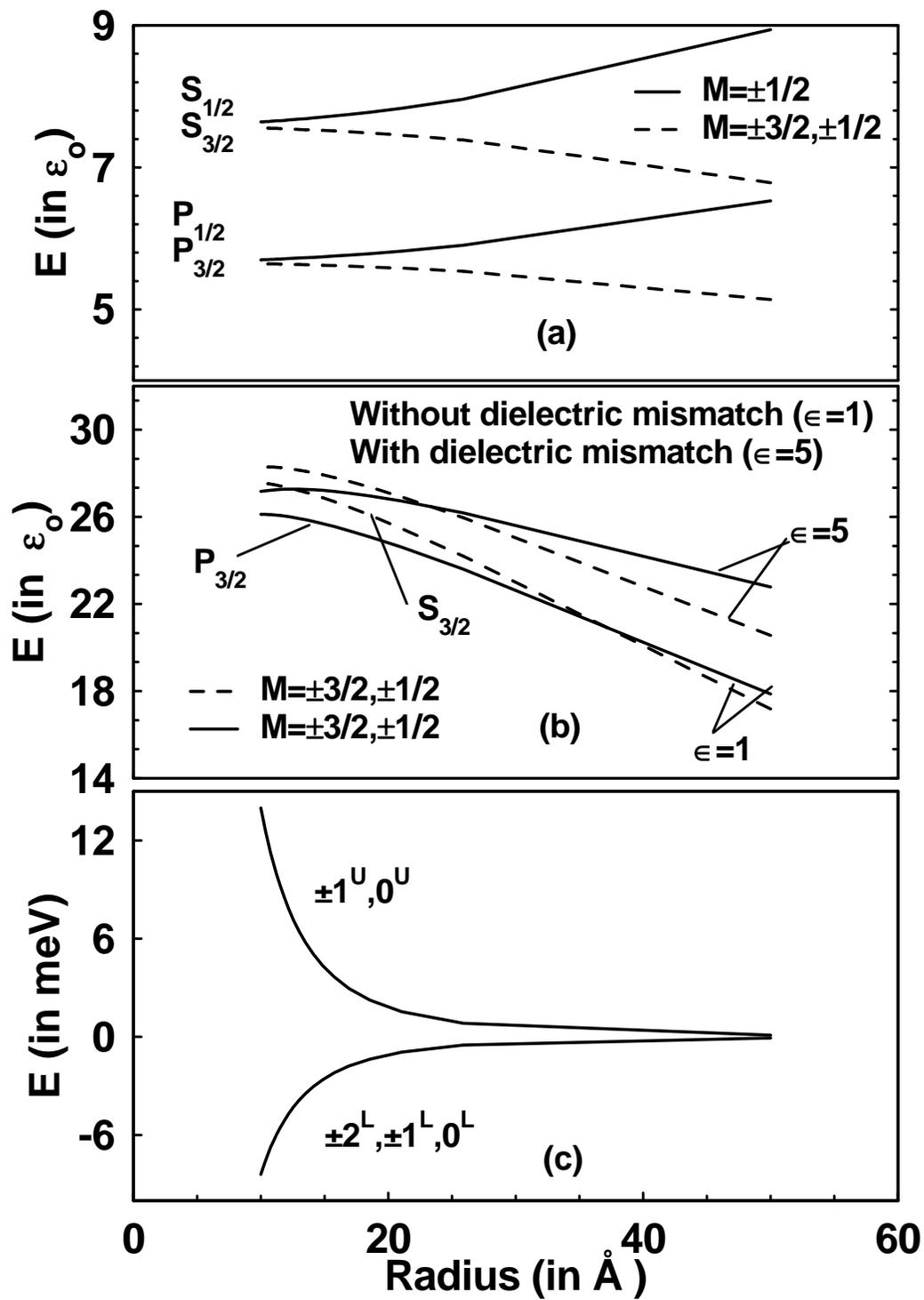
Fig.2:
The energy levels in units of $ \epsilon_{o} $ for z-GaN QDs as a function of dot radius R in the (a) absence and (b) presence of e-h coulomb interaction and dielectric mismatch effects(without exchange). The states are labelled as $S_{\frac{3}{2}}$ etc where the subscripts represent the total angular momentum. (c) eigenvalues including the e-h coulomb and exchange interactions. 

\vspace*{.5cm}
\noindent
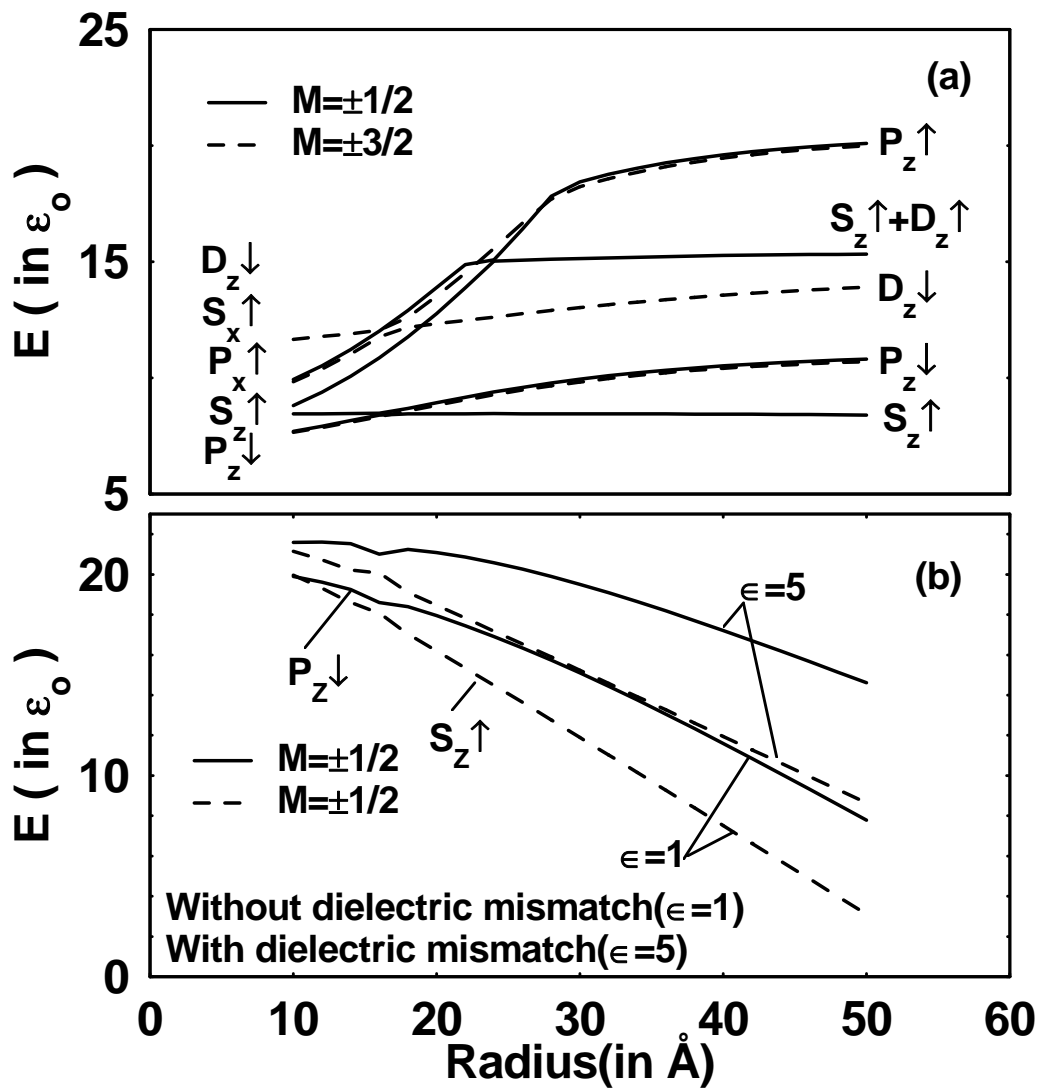
Fig.3:
The energy levels in units of $ \epsilon_{o} $ for w-AlN QDs in the (a) absence and (b) presence of e-h coulomb interaction and dielectric mismatch effects.  

\vspace*{.5cm}
\noindent
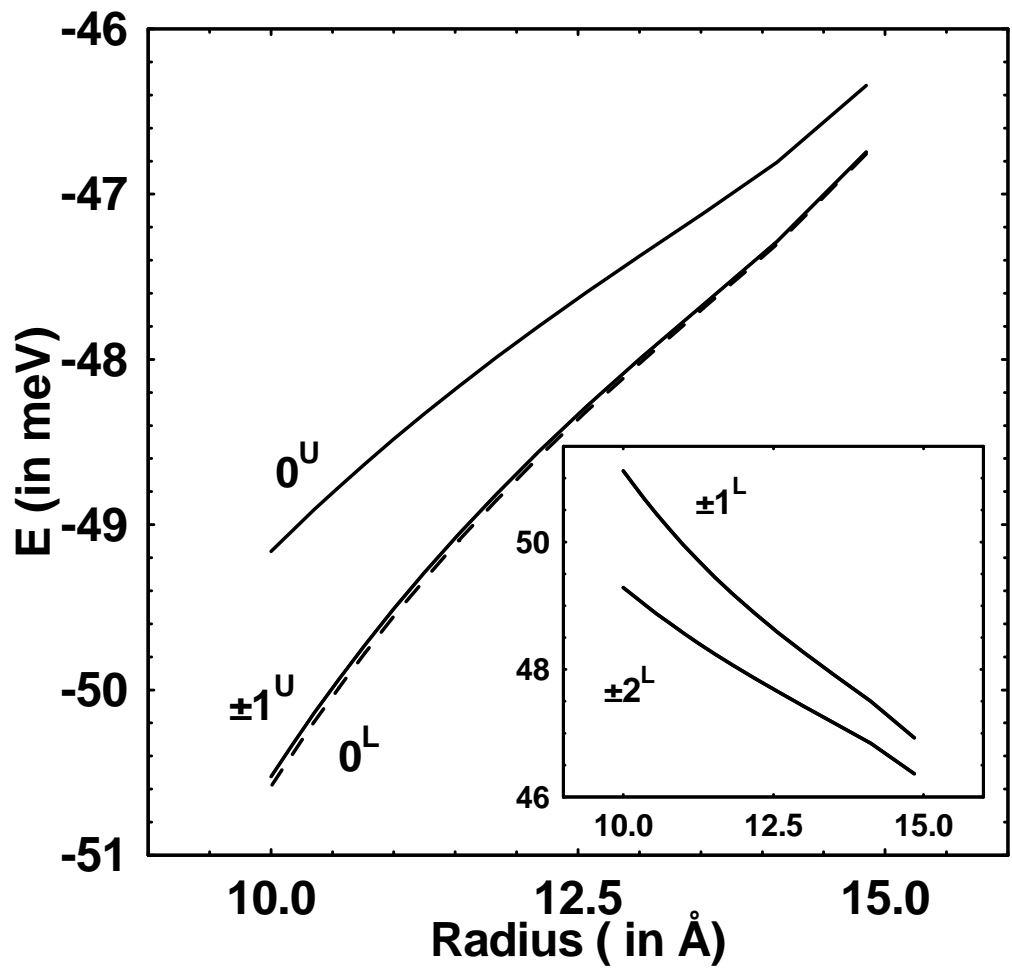
Fig.4:
The energy levels in units of meV for w-AlN QDs after the inclusion of exchange interaction. To show the splitting between the ground state and the first excited state, the ground state $0^{L}$ has been shown as dashed line and the excited states as the solid lines. Few of the excited states has been shown in the inset to clearly show the above splitting. 

\vspace*{.5cm}
\noindent
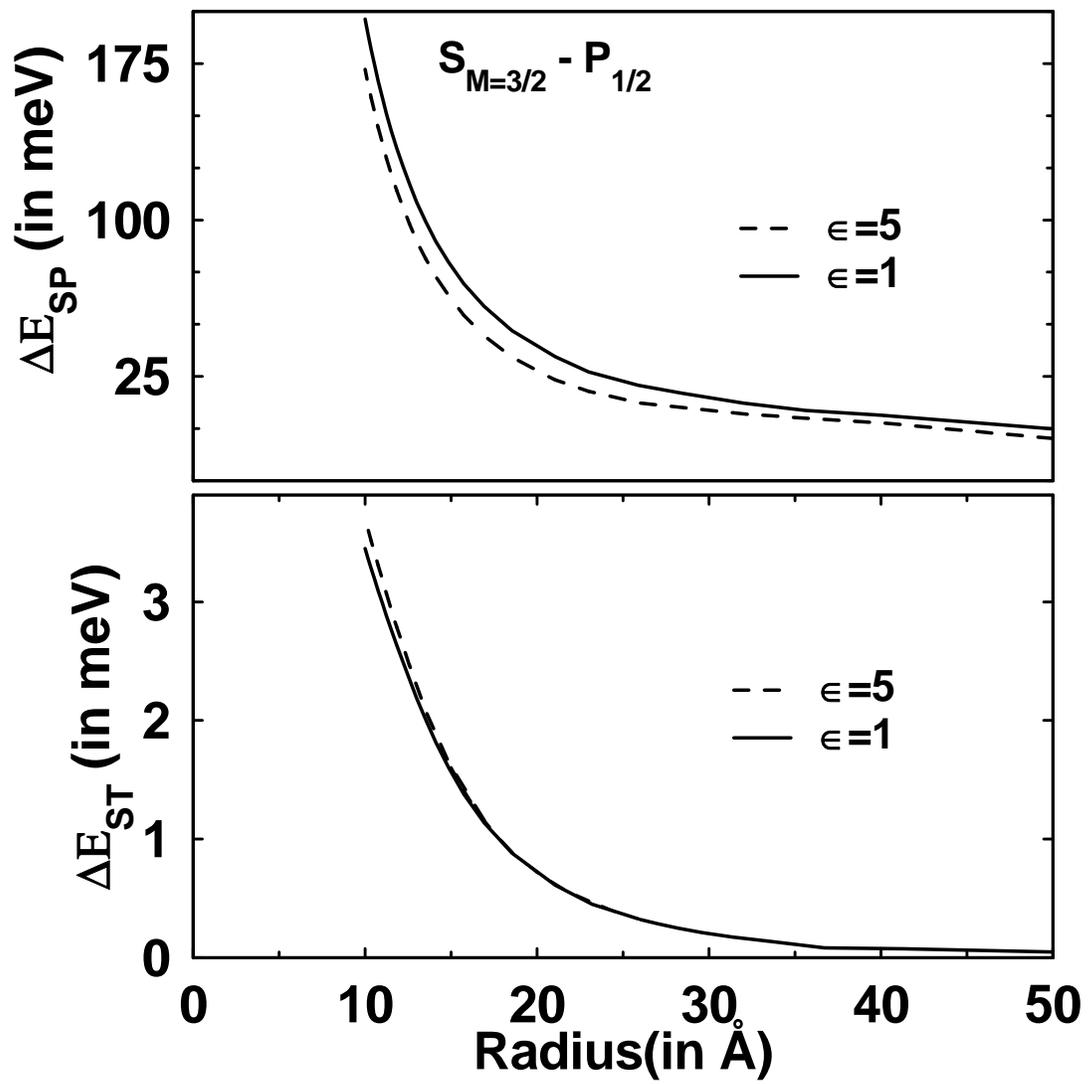
Fig.5:
The energy difference in meV between bright excitonic state and dark excitonic state for w-GaN. (a) Energy difference of S and P states with $M=\frac{3}{2}$ and $\frac{1}{2}$ respectively. (b) The singlet triplet splitting in the absence $\epsilon=\frac{\epsilon_{dot}}{\epsilon_{surrounding}}=1$ and presence $\epsilon=5$ of dielectric mismatch effects.

\vspace*{.5cm}
\noindent
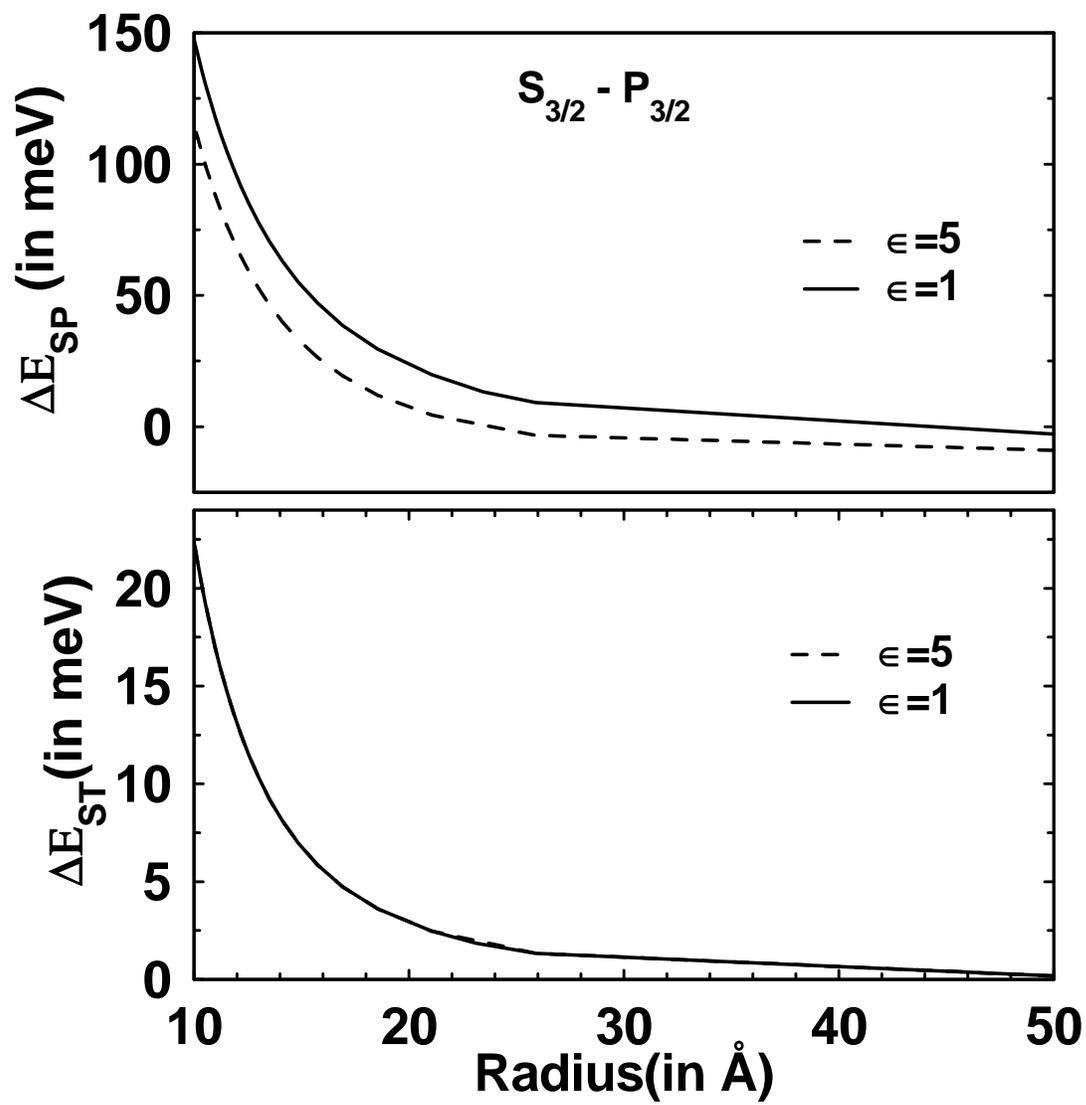
Fig.6:
The energy difference in meV between bright excitonic state and dark excitonic state for z-GaN. (a)Energy differences of S and P states. (b) Singlet triplet splitting in meV in the absence and presence of dielectric mismatch effects.

\vspace*{.5cm}
\noindent
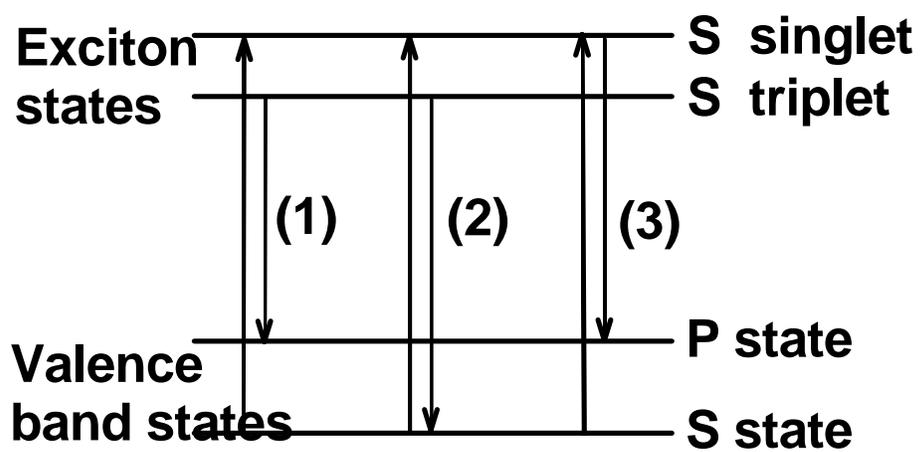
Fig.7:
The excitation-deexcitation processes taking place at the band edges.

\newpage

\begin{center}

{\footnotesize
\begin{tabular}{|l|l|l|l|l|}  \hline
 M &\multicolumn{2}{|c|}{AlN} & \multicolumn{2}{c|}{GaN} \\ \cline{2-5}                          &\multicolumn{1}{|c|}{$R=16$\AA} & \multicolumn{1}{c|}{$R=76$\AA} &                            \multicolumn{1}{c|}{$R=16$\AA} & \multicolumn{1}{c|}{$R=76$\AA} \\ \hline
$\frac{1}{2}$& $\left(P,+1,\downarrow\right)=0.24$ & $\left(S,+0,\uparrow\right)=0.86$  &                    $\left(P,+1,\uparrow\right)=0.50$   & $\left(P,+1,\uparrow\right)=0.54$  \\                 & $\left(P,+0,\downarrow\right)=0.51$ &                                    &
       $\left(P,-1,\uparrow\right)=0.49$   & $\left(P,-1,\uparrow\right)=0.42$\\ \cline{2-5}               & $\left(S,+0,\uparrow\right)=0.63$   & $\left(P,+0,\downarrow\right)=0.85$&                    $\left(P,+1,\downarrow\right)=0.52$  &$\left(S,+1,\downarrow\right)=0.46$ \\                & $\left(P,+0,\downarrow\right)=0.15$ &                                    &
          $\left(P,+0,\downarrow\right)=0.44$ & $\left(D,-1,\downarrow\right)=0.3$\\ \hline   
$\frac{3}{2}$& $\left(P,+1,\uparrow\right)=0.32$   & $\left(P,+0,\uparrow\right)=0.86$  &                    $\left(P,+1,\uparrow\right)=0.52$   & $\left(S,+1,\uparrow\right)=0.78$ \\                  & $\left(P,+0,\uparrow\right)=0.64$   &                                  &
      $\left(P,+0,\uparrow\right)=0.43$   & $\left(D,-1,\uparrow\right)=0.12$ \\ \cline{2-5}  
             & $\left(S,+1,\uparrow\right)=0.78$   & $\left(D,+0,\downarrow\right)=0.88$&                    $\left(S,+1,\uparrow\right)=0.66$   & $\left(P,+1,\uparrow\right)=0.71$ \\                  &                                     &                                  &
          $\left(D,-1,\uparrow\right)=0.28$   &                                   \\ \hline
\end{tabular}
}
\end{center}

\end{document}